\documentclass[letterpaper,11pt]{article}

\usepackage{scrextend}
\usepackage{amsmath,amssymb,epsfig,bbm}
\usepackage[active]{srcltx}
\usepackage[all]{xypic}
\usepackage{MnSymbol}
\usepackage{slashed}
\usepackage{amsthm}

\usepackage{float}

\usepackage{color}

\usepackage{dsfont}

\definecolor{co}{cmyk}{0,0.7,0.3,0}
\definecolor{darkgreen}{cmyk}{1,0,1,.2}
\definecolor{m}{rgb}{1,0.1,1}

\newcommand{\be}{\begin{equation}}
\newcommand{\ba}{\begin{eqnarray}}
\newcommand{\ea}{\end{eqnarray}}
\newcommand{\nn}{\nonumber}

\def\a{\alpha}
\def\b{\beta}

\def\d{\delta}
\def\e{\epsilon}

\def\k{\kappa}

\def\m{\mu}
\def\n{\nu}

\def\p{\pi}

\def\x{\xi}

\def\OO{\Omega}

\def\ca{{\cal A}}
\def\cb{{\cal B}}

\def\cd{{\cal D}}

\def\cf{{\cal F}}

\def\ch{{\cal H}}

\def\cn{{\cal N}}
\def\co{{\cal O}}

\makeatletter
\newcommand{\eqnum}{\refstepcounter{equation}\textup{\tagform@{\theequation}}}
\makeatother

\newcommand{\pa}{\partial}

\newcommand{\cF}{{\cal F}}

\newtheorem*{definition*}{Definition}

\fontfamily{yfrak}

\begin{document}

\vskip 25mm

\begin{center}

{\large\bfseries

Quantum Gravity and the Emergence of Matter

}

\vskip 6ex

Johannes \textsc{Aastrup}$^{a}$\footnote{email: \texttt{aastrup@math.uni-hannover.de}} \&
Jesper M\o ller \textsc{Grimstrup}$^{b}$\footnote{email: \texttt{jesper.grimstrup@gmail.com}}\\ 
\vskip 3ex

$^{a}\,$\textit{Mathematisches Institut, Universit\"at Hannover, \\ Welfengarten 1, 
D-30167 Hannover, Germany.}
\\[3ex]
$^{b}\,$\textit{QHT Gruppen, Copenhagen, Denmark.}
\\[3ex]

{\footnotesize\it This work is financially supported by Ilyas Khan, \\\vspace{-0.1cm}St EdmundÕs College, Cambridge, United Kingdom.}

\end{center}

\vskip 3ex

\begin{abstract}

In this paper we establish the existence of the non-perturbative theory of quantum gravity known as quantum holonomy theory by showing that a Hilbert space representation of the $\mathbf{Q H D} (M) $ algebra, which is an algebra generated by holonomy-diffeomorphisms and by translation operators on an underlying configuration space of Ashtekar connections, exist. We construct operators, which correspond to the Hamiltonian of general relativity and the Dirac Hamiltonian, and show that they give rise to their classical counterparts in a classical limit. We also find that the structure of an almost-commutative spectral triple emerge in the same limit. The Hilbert space representation, that we find, is inherently non-local, which appears to rule out spacial singularities such as the big bang and black hole singularities. Finally, the framework also permits an interpretation in terms of non-perturbative Yang-Mills theory as well as other non-perturbative quantum field theories. This paper is the first of two, where the second paper contains mathematical details and proofs.

\end{abstract}

\newpage
\tableofcontents

\section{Introduction}
\setcounter{footnote}{0}

In this paper and its companion paper \cite{AAA1} we prove the existence of a non-perturbative theory of quantum gravity  known as {\bf quantum holonomy theory} \cite{Aastrup:2015gba,Aastrup:2016ytt} and show that it produces general relativity in a semi-classical limit. In addition to this we show that the structure of a so-called {almost-commutative spectral triple} 
emerge in the same limit -- a result that opens the door to a possible connection to the standard model of particle physics via non-commutative geometry \cite{Connes:1996gi,Chamseddine:2007hz}.\\ 

Quantum holonomy theory is based on an elementary algebra -- called the quantum holonomy-diffeomorphism algebra, denoted $\mathbf{Q H D} (M) $ \cite{Aastrup:2014ppa} -- which is generated first by holonomy diffeomorphisms on a manifold $M$, i.e. translations of tensor degrees of freedom along trajectories of smooth vector fields, and second by certain canonical translation operators on an underlying configuration space $\ca$ of connections over which the holonomy-diffeomorphisms form a non-commutative algebra of functions.

In this paper we first construct a Hilbert space representation of the $\mathbf{Q H D} (M) $ algebra. Since the $\mathbf{Q H D} (M) $ algebra encodes the kinematics of gauge theory as well as quantum gravity formulated in terms of Ashtekar variables  \cite{Ashtekar:1986yd,Ashtekar:1987gu,Barbero:1994ap} it is natural to interpret this Hilbert space in terms of a kinematical sector of a quantum gauge theory and in particular of a theory of quantum gravity. 

A key characteristics of the Hilbert space representation, which we find, is that it does not include local field operators, i.e. operator valued distributions, as we know them from ordinary quantum field theory. 
The reason for this is that the Hilbert space representation is inherently non-local in the sense that the Hilbert space measure weighs different field configurations according to their variation -- i.e. that field configurations, that varies mostly at large scales are assigned higher weight than field configurations, that varies mostly at short scales.
This means in particular that geometrical configurations, that involve spatial singularities, are assigned zero weight in the Hilbert space measure. It is remarkable that this feature appears to rule out the initial big bang singularity as well as the singularities otherwise purported to reside at the centre of black holes. 

This non-locality introduces a scale dependency into the construction so that the Hilbert space representation of the $\mathbf{Q H D} (M) $ algebra comes with two fundamental parameters: one related to quantum gravity via the canonical commutation relations and one related to scale via the just explained non-locality.

The key step in obtaining the Hilbert space representation of the $\mathbf{Q H D} (M) $ algebra is to expand the Ashtekar connection in a orthonormal basis with respect to a Sobolev scalar product and to construct an inner product by integrating over these Sobolev eigenvectors -- this amounts to defining an integration measure on the configuration space of Ashtekar connections. It is this inner product that is sensitive to the variation of the various geometrical variables involved in this particular quantization of gauge theories.  
One immediate  consequence of this approach is that it a priori depends on a background metric since the Sobolev eigenvectors are metric dependent. We propose to interpret this construction as a theory akin to the Ising model, where the representations, that we have found, correspond to a broken phase with a specific choice of background metric.\\

Once we have the Hilbert space representation we move on to construct operators, which correspond to the Hamilton and diffeomorphisms constraint in canonical quantum gravity as well as to the Dirac Hamiltonian, and we show that these operators produce their classical counterparts -- i.e. general relativity\footnote{With the choice of $SU(2)$ as a gauge group we are at the moment dealing with an Euclidean signature.  } and matter -- when evaluated on certain Gaussian states and when a classical limit is taken. It is interesting that it is relatively straight forward to compute these expectation values in closed form without the use of perturbation theory. \\

The original motivation for considering an algebra generated by holonomies was to find an explanation for the appearance of an almost-commutative algebra in the work of Chamseddine and Connes on the standard model of particle physics \cite{Connes:1996gi,Chamseddine:2007hz}. There it was shown that the entire standard model coupled to general relativity can be understood as a purely gravitational theory by adding a matrix factor to the algebra of smooth functions over a four-dimensional manifold $M$ and using the machinery of non-commutative geometry and spectral triples. What remains, then, is to explain: a) the occurrence of the almost commutative algebra and b) the role of quantum field theory in the framework of non-commutative geometry. In \cite{Aastrup:2005yk} we suggested that the answer to these questions should be sought within a framework of pure quantum gravity -- a proposition that lead us to the $\mathbf{Q H D} (M) $ algebra, which sure enough gives rise to an almost-commutative algebra and -- as we find in this paper -- the {\it structure} of an almost-commutative spectral triple within the framework of quantum holonomy theory and in a semi-classical limit. The idea, then, is that quantum field theory emerges as the low-energy limit of a theory of pure quantum gravity. \\

In order to produce a canonical structure, that can serve as a dynamical principle, we write down a Dirac type operator {\it over} the configuration space of connections. This operator involves derivatives in each Sobolev eigenvector and we show that it entails -- via fluctuations of inner automorphisms of the $\mathbf{ H D} (M) $ algebra -- an operator that descends to a spatial Dirac operator on $M$ in a semi-classical limit when evaluated on a Gaussian state. \\

Since the Hilbert space representation of the $\mathbf{Q H D} (M) $ algebra presents us with a general framework of non-perturbative quantum gauge theory it is natural to consider also Yang-Mills theory in this framework. We therefore write down the operator, that corresponds to the Yang-Mills Hamiltonian, and show that it produces its classical counterpart in a semi-classical limit when evaluated on a Gaussian state.  

Also, we consider a generalisation of our framework to field theories such as scalar theories and find that the method of expanding field variables in Sobolev eigenvectors and integrating over each mode provides a viable framework of non-perturbative quantum field theory for a large class of field theories. The quantum field theories, that we find, do not involve local field operators since they permit localisation only up to a scale $\tau_1$, which we tentatively interpret as the Planck scale. The framework breaks down, as expected, in the local limit $\tau_1\rightarrow 0$ where the Sobolev norm descents to the $L^2$-norm.\\

This paper is organised as follows: We start in section 1.1 by giving an outline of the basic idea behind quantum holonomy theory, namely the $\mathbf{Q H D} (M) $ algebra, for then in section 2 to give a proper introduction to the basic algebraic setup. In section 3 we then construct the Hilbert space representation of the $\mathbf{Q H D} (M) $ algebra and compute some simple expectation values on the ground state. We then move on, in section 4, to consider the dynamics, both in terms of a spectral triple type construction over $\ca$ and in terms of operators, that correspond to the constraints of quantum gravity formulated in terms of Ashtekar variables. In section 5 we then discuss background dependency for then, in section 6, to move on to an alternative interpretation in terms of quantum Yang-Mills theory and other field theories. In section 7 we then discuss the absence of singular geometries -- i.e. black holes and big bang -- and finally, in section 8, we show that the structure of an almost-commutative spectral triple emerges from our theory in a semi-classical limit. We end with a discussion in section 9.

\subsection{Outline of the basic idea}

The central idea behind quantum holonomy theory is to build a fundamental theory over an operator algebra generated by holonomies.  A holonomy encodes information about how spinors are parallel transported along paths $\gamma$ in a gauge theory.
\begin{figure}[H]
\begin{center}
\includegraphics[height=4cm, angle=-90]{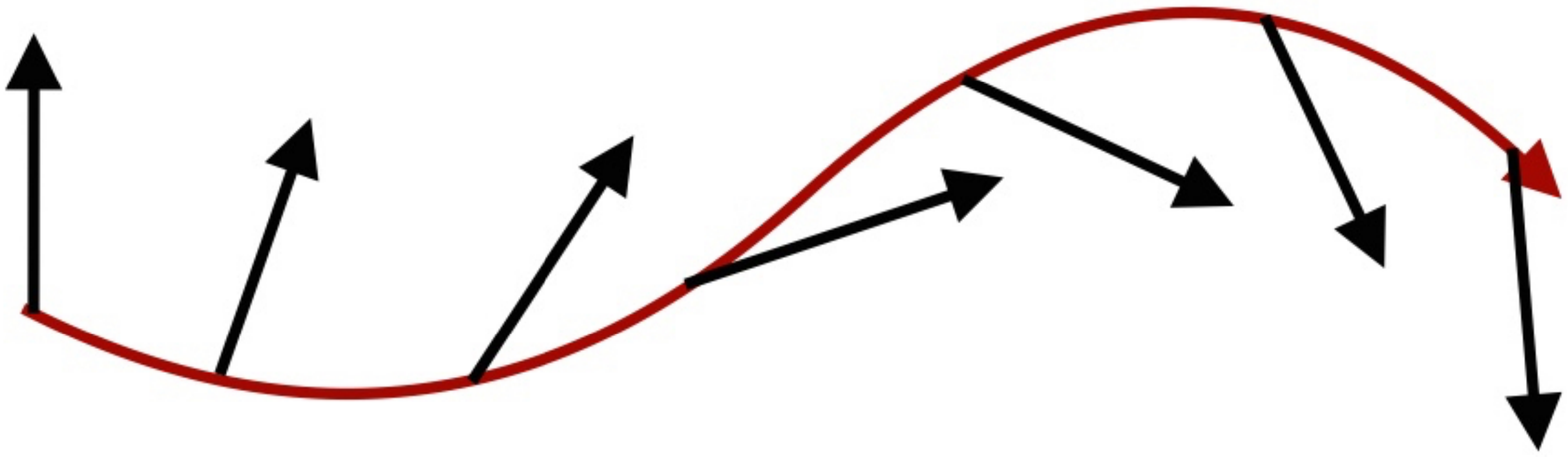}
\end{center}
\caption{Parallel transporting a vector along a path.}
\end{figure}
As an operator the holonomy is best viewed as a map
$$
\ca\ni\nabla \longrightarrow Hol(\gamma,\nabla)\in M_n(\mathbb{C})
$$
from a configuration space $\ca$ of connections into a representation of the corresponding gauge group, i.e. the holonomy tells us how to transform a spinor, which we have moved along the path. 
But the holonomy itself cannot represent a physical quantity since it corresponds to a parallel transport of a single point. All physical theories relevant in current high-energy physics are based on Riemann measures, where single points have zero measure, and thus, if we wish to have an algebra generated by physical quantities, these must be objects with non-vanishing measure\footnote{This point is particularly relevant when discussing quantum gravity, where simple arguments combining quantum mechanics and general relativity show that localisation below the Planck length are operational meaningless. }. To elevate a holonomy operator to a physical quantity we must therefore consider parallel transports of {\it finite volumes} rather than points. This is precisely what a holonomy-diffeomorphism does.
A holonomy-diffeomorphism encodes information about how spinors are parallel transported along the flow of a vector field.
\begin{figure}[H]
\begin{center}
\includegraphics[height=5cm, angle=-90]{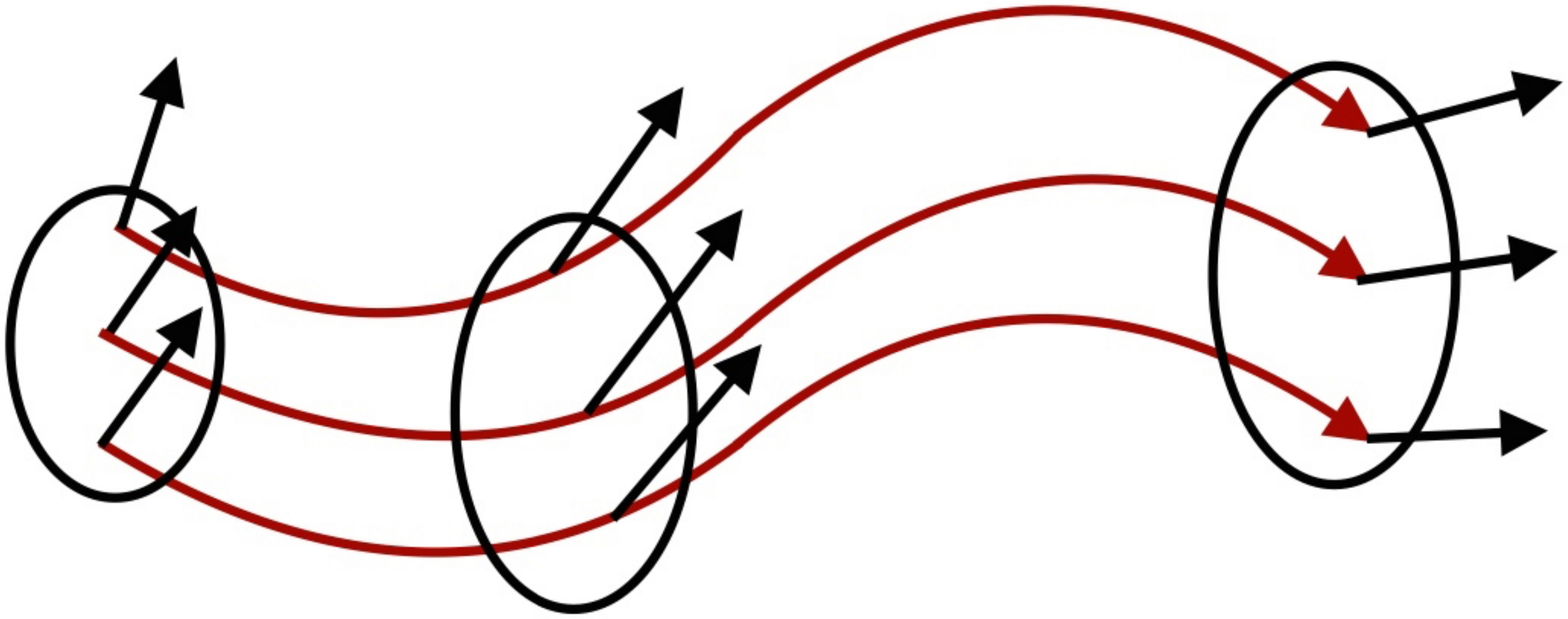}
\end{center}
\caption{Parallel transporting a tensor with local support along the flow of a vector field.}
\end{figure}
A holonomy-diffeomorphism $ e^X$ associated to a vector field $X$ is best understood as a map
$$
\ca\ni\nabla \longrightarrow   e^X_\nabla  \in \cb(L^2(M,S) )
$$
where a connection gives rise to a map $e^X_\nabla$ acting on spinors in a Hilbert space $L^2(M,S)$. This map transports the spinor in its entirety -- i.e. with all of its support -- rather than its value in a single point.

Given the configuration space $\ca$ of gauge connections it is natural to consider translations thereon. Two arbitrary connections $\nabla$ and $\nabla'$ always differ by a one-form $\omega$
$$
\nabla' = \nabla+\omega,
$$
which corresponds to a translation operator $U_\omega$
$$
U_\omega \xi(\nabla) = \xi(\nabla+\omega)
$$
on functions $\xi$ on $\ca$, with the operator identity
\begin{equation}
U_\omega  e^X U_\omega^* (\nabla) = e^X(\nabla-\omega).
\label{larmlarm}
\end{equation}
If we choose the gauge group $SU(2)$ it can be shown \cite{Aastrup:2014ppa} that an infinitesimal version of this operator identity is identical to the canonical commutation relations of quantum gravity formulated in terms of Ashtekar variables\footnote{Note again that with $SU(2)$ we are in fact dealing with the real Ashtekar connection \cite{Barbero:1994ap}, which corresponds to an Euclidean signature.} and to the canonical commutation relations of Yang-Mills theory. This means that the algebra generated by holonomy-diffeomorphisms $e^X$ and translation operators $U_\omega$ -- the $ \mathbf{QHD}(M)$ algebra --  will encode the kinematics of quantum gravity. 
The central idea behind quantum holonomy theory is precisely to obtain a fundamental theory via a Hilbert space representation of the $ \mathbf{QHD}(M)$ algebra.

\section{The Quantum holonomy-diffeomorphism algebra}
\label{firsttask}


We begin with a compact and connected $3$-dimensional manifold $M$ and consider the Hilbert space $L^2(M,S)$ of 2-spinors. Given a vector field $X$ on $M$ with the corresponding flow $t\to \exp_t(X)$ we define the path $\gamma$  
$$\gamma (t)=\exp_{t} (X) (x) $$
running from $x$ to $y=\exp_1 (X)(x)$. Given a $SU(2)$ connection $\nabla$ we then define a map
$$e^X_\nabla :L^2 (M , S) \to L^2 (M ,  S)$$
via the holonomy along the flow of $X$
\begin{equation}
  (e^X_\nabla \xi )(y)=    \hbox{Hol}(\gamma, \nabla) \xi (x)   ,
  \label{chopin1}
 \end{equation}
where $\xi\in L^2(M,S)$ and where $\hbox{Hol}(\gamma, \nabla)$ denotes the holonomy of $\nabla$ along $\gamma$. This map gives rise to an operator valued function on the space $\ca$ of $SU(2)$ connections via
\begin{equation}
\ca \ni \nabla \to e^X_\nabla  ,
\nn
\end{equation}
which we denote by $e^X$. For a function $f\in C^\infty_c (M)$ we get another operator valued function $fe^X$ on $\ca$, which we call a holonomy-diffeomorphisms\footnote{The holonomy-diffeomorphisms, as presented here, are not a priori unitary, but by multiplying with a factor that counters the possible change in volume in (\ref{chopin1}) one can make them unitary, see \cite{AGnew}.}.
The $\mathbf{H D} (M)   $ algebra is then defined as the $C^*$-algebra generated by all holonomy-diffeomorphisms. This algebra, which was first introduced and studied in \cite{Aastrup:2012vq} can be shown to be metric independent. For details on the $\mathbf{H D} (M)   $ algebra we refer the reader to \cite{AGnew}.

Next we let $\mathfrak{su}(2)$ be the Lie-algebra of $SU(2)$.   
A $\mathfrak{su}(2)$ valued one-form induces a transformation of $\ca$, and therefore an operator $U_\omega $ on functions on $\ca$ via   
$$U_\omega (\xi )(\nabla) = \xi (\nabla + \omega) ,$$ 
which satisfy the already mentioned relation 
\begin{equation}
U_\omega fe^X U_\omega^* (\nabla) = fe^X(\nabla-\omega).
\end{equation}
%
%
Infinitesimal translations on $\ca$ are given by 
\begin{equation}
E_\omega  =\frac{d}{dt}U_{  t  \omega}\Big|_{t=0} \;,
\label{soevnloes}
\end{equation}
where we have the relation 
$$
E_{\omega_1+\omega_2}=E_{\omega_1}+E_{\omega_2\;,}
$$
which follows since $U_{(\omega_1+\omega_2 )}=U_{\omega_1}U_{ \omega_2}$. Note that $E_\omega$ is skew-adjoint.
We finally define the $\mathbf{QHD}(M)$ algebra as the algebra generated by $\mathbf{HD}(M)$ and by translations $U_{\omega}$.

For more details on the $\mathbf{HD}(M)$ and $\mathbf{QHD}(M)$ algebras we refer the reader to these publications \cite{Aastrup:2015gba,Aastrup:2016ytt,Aastrup:2014ppa,Aastrup:2016caz}.

\subsection{Relation to canonical quantum gravity }

The $\mathbf{QHD}(M)$ algebra is closely related to canonical quantum gravity formulated in terms of Ashtekar variables as well as to Yang-Mills theory. This relation, which was first discussed in \cite{Aastrup:2014ppa}, is best seen by computing the commutator between $E_\omega$ and an element $f e^X$ of $\mathbf{HD}(M)$:
\begin{equation}
[E_\omega, f e^X](\nabla) \xi(y) = \int_\gamma dt  Hol(\gamma_{<t},\nabla) \omega(\dot{\gamma})(t) Hol(\gamma_{>t},\nabla) \xi(x), 
\label{comnu}
\end{equation}
where $\xi\in L^2(M,S)$ and where $\gamma: [0,1] \rightarrow M$ is the path generated by the vector field $X$ with $\gamma(0)=y$ and $\gamma(1)=x$. Also, $\gamma_{<t}$ is the section $\gamma_{<t}: [0,t]\rightarrow M$ and likewise $\gamma_{>t}: [t,1]\rightarrow M$.
This commutator reproduces the structure of the corresponding Poisson bracket in canonical quantum gravity between the flux $F_S$ of the inverse triad field $E^\m_i $ and a holonomy of the Ashtekar connection $A_\n^j $
$$
\{F^a_S, Hol(\gamma,A)\}_{\mbox{\tiny P.B.}} = \pm Hol (\gamma_1,A) \sigma^a Hol(\gamma_2,A)
$$
where $\gamma=\gamma_1\circ\gamma_2$ and where $S$ is a surface that intersects $\gamma$ at the point where the Pauli matrix $\sigma$ is inserted. To see the correspondence to the Ashtekar variables themselves we can also define the infinitesimal holonomy-diffeomorphism
\begin{equation}
{\nabla}_X = \frac{d}{dt} e^{tX}\Big|_{t=0} 
\label{covvar}
\end{equation}
and consider its commutator with $E_\omega$. This was done in \cite{Aastrup:2014ppa}, where it was seen to reproduce the structure of the canonical commutation relations
\begin{equation}
\{ E^\m_i (x), A_\n^j (y) \}_{\mbox{\tiny P.B.}} = \d(x-y) \d_i^j \d_\n^\m .
\label{fix}
\end{equation}
This Poisson bracket is of course identical to that of a Yang-Mills theory with an appropriate choice of gauge group. 
We refer the reader to \cite{Aastrup:2016ytt} for more details.

\section{A Hilbert space representation of $\mathbf{QHD}(M)$}

In this section we construct a Hilbert space representation for the $\mathbf{QHD}(M)$ algebra.

\subsection{The Hilbert space }

We begin by constructing a Hilbert space, which can be understood as $L^2(\ca)$ 
where $\ca$ is the space of smooth $SU(2)$ connections.
To construct this Hilbert space we first let 
$\langle \cdot\vert\cdot\rangle_{\mbox{\tiny sob}}$ denote the Sobolev norm on $\OO^1(M\otimes\mathfrak{su}(2))$, which has the form
\begin{equation}
\langle \omega_1\vert\omega_2\rangle_{\mbox{\tiny sob}}
:=
\int_M dx \mbox{Tr}_{M_2} \big( (1+ \tau_1\Delta^{\sigma})\omega_1  , (1+  \tau_1\Delta^{\sigma})\omega_2  \big)_{T_x^*M}
\label{sob}
\end{equation}
where the Hodge-Laplace operator $\Delta$ and the  inner product  $(,)_{T_x^*M}$ on $T_x^*M$ depend on a metric g and where $\tau_1$ and $\sigma$ are positive constants with\footnote{We could here also choose higher exponents  or even the exponential of the Hodge-Laplace operator, leading to viable constructions. In the present paper we consider, however, merely the lowest possible exponent of the Hodge-Laplace operator.} $\sigma > \frac{5}{4}$. Note that the Sobolev norm descents to an $L^2$ norm in the limit $\tau_1\rightarrow 0$
\begin{equation}
\langle \omega_1\vert\omega_2\rangle_{\mbox{\tiny sob}} \stackrel{\tau_1\rightarrow 0}{\longrightarrow}  \langle \omega_1\vert\omega_2\rangle_{L^2} .
\label{soblimit}
\end{equation}
Denote by $\{\phi_i\}_{i\in\mathbb{N}}$ an orthonormal basis of $\OO^1(M\otimes\mathfrak{su}(2))$, where $\phi_i$ are eigenvectors of the Hodge-Laplace operator, i.e. $\Delta \phi_i = \lambda_i^2 \phi_i$, and let  $\{\xi_i\}_{i\in\mathbb{N}}$ be an orthonormal basis of $\OO^1(M\otimes\mathfrak{su}(2))$ with respect to the scalar product (\ref{sob}) of the form
\begin{equation}
\xi_i = \frac{\phi_i}{1+\tau_1 \lambda_i^{2\sigma}}.
\label{LTF}
\end{equation}
This means that the $L^2$-norm of $\xi_i$ is $\tfrac{1}{1+\tau_1 \lambda_i^{2\sigma}}$. 
Moreover, the Sobolev eigenvectors $\{\xi_i\}_{i\in\mathbb{N}}$ satisfy the relation
\begin{eqnarray}
\sum_{i=1}^\infty \Vert \xi_i \Vert^2_{\mbox{\tiny sup}} < \infty
\label{houston}
\end{eqnarray}
where $\Vert \cdot \Vert^2_{\mbox{\tiny sup}}$ is the supremum norm. For details we refer the reader to \cite{AAA1}. Also, we choose the labelling of $\{\xi_i\}$ to follow the increasing size of eigenvalues $\lambda_i$, i.e. $\lambda_1 \leq \lambda_2 \leq \lambda_3 ...$, etc.

We are now ready to construct the Hilbert space $L^2(\ca)$. Let $\eta,\zeta$ be two functions on $\ca$ on the form $\eta(x_1\xi_1\ldots x_n\xi_n)$ and $\zeta(x_1\xi_1\ldots x_n\xi_n)$ and define the inner product
\begin{equation}
\langle \eta \vert \zeta \rangle_{\ca_n} := \int_{\mathds{R}^n}  \overline{\eta(  x_1\xi_1 + \ldots +x_n \xi_n )} \zeta ( x_1\xi_1 + \ldots + x_n \xi_n ) dx_1\ldots dx_n,
\label{Richter}
\end{equation}
which gives rise to a Hilbert space $L^2(\ca_n)$ over the space spanned by the first $n$ Sobolev eigenvectors $\xi_i$. 
Next we note that there is a Hilbert space embedding 
$$
\varphi_n : L^2(\ca_n)\rightarrow L^2(\ca_{n+1})
$$
given by
\begin{equation}
\varphi_n (\eta)(  x_1\xi_1 + \ldots +x_n \xi_n + x_{n+1} \xi_{n+1}  ) =   \eta(  x_1\xi_1 + \ldots +  x_n\xi_n   ) \frac{1}{\sqrt[4]{\pi}}e^{- \frac{x_{n+1}^2}{2\tau_2}},
\label{indian}
\end{equation}
which leads us to construct $L^2(\ca)$ as the inductive limit
$$
L^2(\ca_1) \stackrel{\varphi_1}{\longrightarrow} L^2(\ca_2) \stackrel{\varphi_2}{\longrightarrow} \ldots   \stackrel{\varphi_n}{\longrightarrow} L^2(\ca_{n+1}) \stackrel{\varphi_{n+1}}{\longrightarrow} \ldots ,
$$
which is separable. We denote the inner product on $L^2(\ca)$ by $\langle \cdot\vert\cdot\rangle_\ca$ and shall occasionally use the notation
\begin{equation}
\langle \eta \vert\zeta\rangle_\ca = \int_\ca d\omega \overline{\eta(\omega)} \zeta(\omega).
\label{Richter2}
\end{equation}
The Hilbert space, in which we find a representation of the $\mathbf{QHD}(M)$ algebra, is then constructed as 
$$
\ch = L^2(\ca) \otimes L^2(M,S)
$$
i.e. it consist of tensor products between functions on $\ca$ and spinors on $M$. We denote the inner product on $\ch$ by $\langle \cdot\vert\cdot\rangle_\ch$. For details on the construction of $L^2(\ca)$ and $\ch$ we refer the reader to \cite{AAA1}.

Note that the inner product written in (\ref{Richter2})  in fact depends on the choice of a fixed connection $\nabla_0\in\ca$ via $\langle \eta \vert\zeta\rangle_\ca = \int_\ca d\omega \overline{\eta(\nabla_0+\omega)} \zeta(\nabla_0+\omega)$. We choose to set $\nabla_0=0$ in the following.

Before we continue let us also point out that much of the ensuing analysis will be based on the fact that the map
\begin{equation}
\ca \ni \omega=\sum_i x_i \xi_i \rightarrow (x_1,x_2,\ldots)\in \mathds{R}^\infty
\label{mapp}
\end{equation}
embeds the configuration space $\ca$ in $\mathds{R}^\infty$ and likewise $L^2(\ca)$ in $L^2(\mathds{R}^\infty)$ (understood as projective and inductive limits respectively). Once this correspondence is established the analysis essentially boils down to defining quantum mechanics on the projective limit $\mathds{R}^\infty$.

\subsection{The operators}

We are now ready to write down the representation. 
Given a smooth one-form $\chi\in\OO^1(M,\mathfrak{su}(2))$ we write $\chi =\sum a_i \xi_i$. The operator $U_\chi$  acts by translation in $L^2(\ca)$, i.e. 
\begin{eqnarray}
U_{\chi}(\eta) (\omega)&=&U_{\chi}(\eta) (x_1 \xi_1+x_2 \xi_2+ \ldots)
\nn\\
&=&  \eta ( (x_1+a_1)\xi_1+(x_2+a_2)\xi_2+ \ldots)  . 
\label{rep1}
\end{eqnarray}
with $\eta\in L^2(\ca)$. Next, we let $f e^X\in \mathbf{HD}(M)$ be a holonomy-diffeomorphism and $\Psi(\omega,x)=\eta(\omega)\otimes \psi(x)\in\ch$. We write
\begin{equation}
f e^X \Psi(\omega,y) =  f(x) \eta(\omega) Hol(\gamma, \omega) \psi(x)  
\label{rep2}
\end{equation}
where $\gamma$ is the path generated by the vector field $X$ with $y=\exp_1(X)(x)$. In \cite{AAA1} we prove that (\ref{rep1}) and (\ref{rep2}) give rise to a Hilbert space representation of the $\mathbf{QHD}(M)$ algebra.

To see how this representation works let us first consider a state $\eta_{\bf A}$, which is localised over a smooth connection\footnote{Note that we have chosen to place the Gauss factors in the states, such as in (\ref{groundstate}), and not in the measure of $L^2(\ca)$. The reason for this is that the adjoint of $U_\omega$ has a simpler form this way (else it would involve the conjugate of the Gauss factors). The correct choice, however, is to place the Gauss factors in the measure. When we do this the state (\ref{groundstate}) is simply the identity. This issue becomes important when we analyse the projective and inductive limits. The reason for this is that the measure is in fact not a Lebesgue measure. For details see \cite{AAA1}.}  ${\bf A}$ in $\ca$, i.e. 
\begin{eqnarray}
\eta_{\bf A} (x_1 \xi_1+\ldots x_n \xi_n+ \ldots) \hspace{-3cm}\nn\\
&=& \frac{1}{\sqrt[4]{\tau_2 \pi}}e^{-\frac{(x_{1}-a_1)^2}{2\tau_2}}\cdot \frac{1}{\sqrt[4]{\tau_2\pi}}e^{-\frac{(x_{2}-a_2)^2}{2\tau_2}}\cdots \frac{1}{\sqrt[4]{\tau_2\pi}}e^{-\frac{(x_{n}-a_n)^2}{2\tau_2}} \cdots ,
\label{groundstate}
\end{eqnarray}
where we use the expansion ${\bf A}=\sum a_i \xi_i$ and where $\tau_2$ is a constant. We can also write this as
$$
\eta_{\bf A}(\omega)= \cn^{-\frac{1}{2}} \exp\left( - \frac{1}{2\tau_2} \Vert \omega - {\bf A} \Vert^2_{\mbox{\tiny sob}}\right),
$$
where $\cn$ is the appropriate normalisation. We then find
\begin{equation}
\langle \eta_{\bf A}\vert U_{\chi }\vert \eta_{\bf A} \rangle_{\ca} = e^{-\frac{1}{4\tau_2}\sum_{i=1}^\infty   (b_{i}-a_i)^2}=e^{-\frac{1}{4\tau_2} \| \chi -  {\bf A} \|_{\mbox{\tiny sob}}^2}.
\end{equation}
where $\chi =\sum b_i \xi_i$. This means that the transition function between two different points ${\bf A}$ and ${\bf A}'$ in the configuration space $\ca$ will depend on the Sobolev norm of their difference $\d={\bf A} - {\bf A}'$. If $\d$ varies only on a large scale, i.e. if it has a {\it small} Sobolev norm, then the transition function will be relatively {\it large} compared to the opposite situation, where $\d$ varies on a short scale. This shows that the entire representation is scale dependent and non-local. In particular, if $\chi$ is a one-form localised in a single point its Sobolev norm will be infinite\footnote{this is best seen by the fact that the Sobolev norm dominates the supremum norm when $\sigma$ is chosen sufficiently large \cite{Nirenberg}.} and the corresponding transition function will vanish. We shall comment further on this in section \ref{blackholes}.

Next, let us consider the expectation value of a holonomy-diffeomorphism. To this end we let $\psi$ be 
 a spinor in $L^2(M,S)$ and write
\begin{eqnarray}
\langle \eta_{\bf A} \otimes \psi \vert f e^X \vert\eta_{\bf A} \otimes\psi\rangle_\ch  
 \hspace{-3cm}&&\nn\\
&=&
 \lim_{n\rightarrow\infty} \frac{1}{\left(\sqrt{\tau_2\pi}\right)^{n}}  \int_{\mathds{R}^n} dx_1\ldots dx_n  \big( \overline{\psi}, f Hol(\gamma,  \sum_{i=1}^n x_i \xi_i) \psi\big) e^{- \sum \frac{(x_i-a_i)^2}{\tau_2} }
\nn\\
&=&
\int_\ca d\omega  \left( \overline{\psi}, f Hol(\gamma,  \omega) \psi \right) e^{-  \frac{1}{\tau_2} \Vert   \omega- {\bf A}\Vert^2_{\mbox{\tiny sob}} }
\nn\\
&=&
\int_\ca d\omega  \left( \overline{\psi}, f Hol(\gamma, {\bf A}+ \omega) \psi \right) e^{- \frac{1}{\tau_2} \Vert  \omega\Vert^2_{\mbox{\tiny sob}} },
\label{pathintegral}
\end{eqnarray}
where $(,)$ is the inner product on $L^2(M,S)$. Likewise we can write down the expectation value of the covariant derivative $\nabla_X$
\begin{eqnarray}
\langle \eta_{\bf A} \otimes \psi \vert \nabla_X \vert\eta_{\bf A} \otimes\psi\rangle_\ch  
=
\int_\ca d\omega  \left( \overline{\psi},  (d+ {\bf A}+ \omega)(X) \psi \right) e^{- \frac{1}{\tau_2} \Vert  \omega\Vert^2_{\mbox{\tiny sob}} }.
\label{pathintegral2}
\end{eqnarray}
Note that these equations have the form of path integrals over the configuration space of gauge connections, where the Sobolev norm provides the weight. Once more we see the scale dependency of this representation of the $\mathbf{QHD}(M)$ algebra, where the path integral assigns larger weight to those parts of the configuration space $\ca$, which varies at large scales -- i.e. with {\it small} Sobolev norm -- and less weight to those parts, which varies at short scales.  \\

A key element in the proof that (\ref{rep1}) and (\ref{rep2}) give rise to a Hilbert space representation of the $\mathbf{QHD}(M)$ algebra is to see that the expectation value (\ref{pathintegral}) of a holonomy-diffeomorphism on a Gaussian state exist. To see why this is the case let us consider the $U(1)$ case, where the holonomy is Abelian and thus
$$
Hol(\gamma, {\bf A} + {\bf B}) =Hol(\gamma, {\bf A})Hol(\gamma, {\bf B}) .
$$
We then have the estimate
$$
\left|    \int_0^{t_0} dt \xi_i (\dot{\gamma}) (\gamma(t))      \right| \leq  t_0\|  \xi_i \|_{\sup},
$$
where $t_0$ is the length of $\gamma$ and $\|  \cdot \|_{\sup}$ the supremum norm. With this we can estimate the integral
$$ 
\frac{1}{\sqrt{\pi}} \int_{-\infty}^\infty dx Hol(\gamma, {\bf A}+ x\xi_i) e^{-x^2}    =\frac{ Hol(\gamma , {\bf A})}{\sqrt{\pi}} \int_{-\infty}^\infty  dx e^{ i x \int_0^{t_0} dt \xi_i (\dot{\gamma}) (\gamma(t))  } e^{-x^2}     ,
$$
with
$$
\frac{1}{\sqrt{\pi}}\int_{-\infty}^\infty  dx e^{i x  \int_0^{t_0} dt \xi_i (\dot{\gamma}) (\gamma(t))  } e^{-x^2}    \geq    e^{-t^2_0 \| \xi_i \|_{sup}^2} .
$$
Here the Gaussian factor $e^{-x^2}$ comes from the state in $L^2(\ca)$. The convergence of the expectation value of a holonomy-diffeomorphism can then be determined by knowing that the supremum norm of the Laplace eigenfunctions $\phi_i$ are bounded by $\lambda_i$ \cite{Grieser} and by setting $\sigma$ in (\ref{sob}) to be greater than $\frac{5}{4} $. For details we refer to \cite{AAA1}.\\

Another important element in the proof is to establish strong continuity w.r.t. the manifold $M$ in the sense of the Sobolev norm between paths:
$$
\| \gamma_1 -\gamma_2\|_{\mbox{\tiny sob}} := \sup_{t\in [0,1]} ( \|\gamma_1(t)-\gamma_2(t)\| +\|\gamma_1'(t)-\gamma_2'(t) \|).
$$
In \cite{AAA1} we prove that
if a sequence of paths $\{\gamma_k\}$ approaches the path $\gamma$
$$
\| \gamma_k -\gamma \|_{\mbox{\tiny sob}} \to 0
$$ 
then so do the corresponding expectation values in $\ch$ of the holonomy-diffeomorphisms
$$
\langle \eta (\omega) \vert Hol(\gamma_k,\omega)\vert \eta(\omega) \rangle_\ch  \to \langle \eta(\omega)\vert Hol(\gamma,\omega)\vert \eta(\omega) \rangle_{\ch\otimes \mathbb{C}^2} ,
$$
with $\eta\in\ch\otimes \mathbb{C}^2$, which implies that we have strong continuity. This result shows that this construction encodes information about the differential structure of the manifold $M$. This is in stark contrast to the measures used in for example loop quantum gravity \cite{Ashtekar:2004eh} and previously by ourselves \cite{Aastrup:2008wa,Aastrup:2008wb}, where the smooth connections have zero measure. \\

Finally, it is worth noting that (\ref{pathintegral}) exist in any dimension as long as we choose $\sigma$ in (\ref{sob}) large enough. The reason for this is precisely its scale dependency, where the Sobolev norm provides what might be called a soft cut-off.

\subsection{The triad field}
\label{sectiontriadfield}


Let us again consider the state $\eta_{\bf A}  \in L^2(\ca)$ from (\ref{groundstate}) and let us modify it by adding a phase
\begin{eqnarray}
\eta_{  ({\bf A, E}) } (\nabla + x_1\xi_1 + \ldots x_n\xi_n + \ldots) \hspace{-5.5cm}&&\nn\\
&&=  \frac{1}{\sqrt[4]{\tau_2\pi}}e^{-\frac{(x_{1}-a_1)^2}{2 \tau_2 }+\frac{ i x_1b_1}{\tau_2}}\cdot \frac{1}{\sqrt[4]{\tau_2\pi}}e^{-\frac{(x_{2}-a_2)^2}{2 \tau_2 }+ \frac{ i x_2b_2}{\tau_2}}\cdots \frac{1}{\sqrt[4]{\tau_2\pi}}e^{-\frac{(x_{n}-a_n)^2}{2 \tau_2 }+ \frac{i x_n b_n}{\tau_2}} \cdots .\nn\\
\label{skudiskoven}
\end{eqnarray}
This phase, which is parametrised by the sequence $\{b_i\}$, can be interpreted in two different ways. First, if we let ${\bf e}= \sum b_i \xi_i$ be a one-form that takes values in $\mathfrak{su}(2)$, i.e. with $b_i=\langle {\bf e} \vert \xi_i \rangle_{\mbox{\tiny sob}} = (1+\tau_1\lambda_i^{2\sigma}){\bf e}_i $ where ${\bf e}=\sum_i {\bf e}_i\phi_i$, then we can rewrite (\ref{skudiskoven}) as
\begin{eqnarray}
\eta_{  ({\bf A, E}) } (\omega)
&=& \cn^{-\frac{1}{2}} \exp \left(- \frac{1}{2\tau_2} \Vert \omega-{\bf A} \Vert^2_{\mbox{\tiny sob}} + \frac{i}{\tau_2} \langle {\bf e} \vert \omega \rangle_{\mbox{\tiny sob}}   \right)
\end{eqnarray}
where the label ${\bf E}$ is a smooth {\it inverse} triad field given by ${\bf E}= g({\bf e},\cdot)$, where $g$ is again the metric used to define the Sobolev norm in (\ref{sob}). 
With this we find 
\begin{equation}
\langle \eta_{  ({\bf A, E}) }  \vert U_{\chi} \vert \eta_{  ({\bf A, E}) } \rangle_\ca = e^{ -\frac{1}{4  \tau_2 } \Vert \chi \Vert^2_{\mbox{\tiny sob}} -\frac{i}{\tau_2} \langle {\bf e} \vert \chi \rangle_{\mbox{\tiny sob}}   }
\label{KIMTRUMP}
\end{equation}
as well as
\begin{equation}
\langle \eta_{  ({\bf A, E}) }  \vert i \tau_2 E_\chi \vert \eta_{  ({\bf A, E}) }  \rangle_\ca   =   \langle {\bf e} \vert \chi \rangle_{\mbox{\tiny sob}}  .
\label{datasikkerhed}
\end{equation}
With the definition of the Sobolev norm (\ref{sob}) we have
$$
\langle {\bf e} \vert \chi \rangle_{\mbox{\tiny sob}}  = \int_M \mbox{Tr}_{M_2} {\bf E} (\chi) + \co(\tau_1).
$$
This result aligns well with our previous interpretation of $E_\chi$ as an operator related to the (inverse) triad field.
It is, however, important to note that $E_\chi$ is only a well defined operator in $L^2(\ca)$ whenever $\Vert \chi \Vert^2_{\mbox{\tiny sob}}<\infty$. This implies that we {\it cannot} define the local operator
\begin{equation}
\hat{E}^\m_i (x) := i \tau_2 E_{\d(x) dx^\m\sigma^i} \qquad \mbox{(!)}
\label{localop}
\end{equation}
in $L^2(\ca)$ since the one-form $\d(x) dx^\m\sigma^i$ has infinite Sobolev norm due to the delta function. 
This shows that quantum holonomy theory does not permit operator valued distributions as we know them from ordinary quantum field theory.

There exist an alternative interpretation of the parametrisation $\{b_i\}$ of the phases in (\ref{skudiskoven}). If we instead let ${\bf E}$ be a densitised inverse triad field\footnote{In fact, this is precisely the original Ashtekar variable conjugate to the connection.} and let
\begin{equation}
b_i = \int_M \mbox{Tr}_{M_2}   {\bf E}(\xi_i)
\label{thulesendahl}
\end{equation}
which is independent of the metric $g$, then we obtain instead 
$$
\sum_i x_i b_i = \int_M \mbox{Tr}_{M_2}  {\bf E} (\omega)
$$
with $\omega=\sum_i x_i \xi_i$. Using (\ref{LTF}) we can rewrite (\ref{thulesendahl}) as
$$
b_i = \frac{\bf e_i}{1+\tau_1\lambda_i^{2\sigma}} 
$$  
where ${\bf e}= g({\bf E,\cdot}) = \sum {\bf e}_i \phi_i $ is once again a one-form obtained from ${\bf E}$ where $\{\phi_i\}_{i\in\mathbb{N}}$ is still an orthonormal basis of eigenvectors of the Hodge-Laplace operator.
With this interpretation the phase in (\ref{KIMTRUMP}) would not involve the Sobolev norm and equation (\ref{datasikkerhed}) would instead have the form
\begin{equation}
\langle \eta_{  ({\bf A, E}) }  \vert i \tau_2 E_\chi \vert \eta_{  ({\bf A, E}) }  \rangle_\ca   =    \int_M \mbox{Tr}_{M_2}  {\bf E} (\chi) \qquad\mbox{(alt. interpretation)} 
\label{datasikkerhed2}
\end{equation}

The difference between these two interpretations of the phase essentially boils down to whether or not the background metric, that is used to define the Sobolev norm, is included into the data ${\bf E}$, and whether or not the phase involves the Sobolev norm or the $L^2$-norm. In the following we shall mostly use the first interpretation. We shall make it clear when this distinction is important.\\

Let us finally consider Fourier transformations in $L^2(\ca)$. If $\eta(\omega)$ is a state in $L^2(\ca)$ then we write its Fourier transformation as
$$
\tilde{\eta} (\chi) = \int_\ca d\omega \eta(\omega) e^{ - \frac{ i}{\tau_2} \langle \omega \vert \chi \rangle_{\mbox{\tiny sob}}    }.
$$
Let us for instance consider the state $ \eta_{  ({\bf A, E}) } (\omega) $ where we find
$$
\tilde{\eta}_{  ({\bf A, E}) } (\chi) = \cn^{\frac{1}{2}} \exp\left( - \frac{2}{\tau_2}  \Vert {\bf e} - \chi \Vert^2_{\mbox{\tiny sob}}  + \frac{i}{\tau_2} \langle {\bf A}\vert {\bf e}-\chi\rangle_{\mbox{\tiny sob}}      \right),
$$
which is a Gaussian centred over the point ${\bf e}$ and where ${\bf A}$ appears in a phase, as one might have expected. 
This shows that the gauge field ${\bf A}$ and the triad field ${\bf e}$ are dual entities related via a Fourier transform. 


\subsection{On a semi-classical limit}

Let us briefly consider  the limit $\tau_2\rightarrow 0$ of the expectation values of the holonomy and translation operators that we have seen so far.
We consider therefore the state $\eta_{  ({\bf A, E}) } \otimes\psi \in \ch$ in (\ref{skudiskoven}). Using (\ref{pathintegral}) we first find
\begin{eqnarray}
\lim_{\tau_2\rightarrow 0}\langle \eta_{  ({\bf A, E}) }  \otimes \psi \vert f e^X \vert\eta_{  ({\bf A, E}) } \otimes\psi\rangle_\ch  
=  \int_M \overline{\psi} (x)f(x)Hol(\gamma, {\bf A}) \psi (x),
\end{eqnarray}
as well as
\begin{eqnarray}
\lim_{\tau_2\rightarrow 0}\langle \eta_{  ({\bf A, E}) }  \otimes \psi \vert\nabla_X \vert\eta_{  ({\bf A, E}) } \otimes\psi\rangle_\ch  
=  \int_M \overline{\psi} (x)  (d + {\bf A}) (X)   \psi (x).
\end{eqnarray}
Likewise, we find that the transition function vanishes
$$
\lim_{\tau_2\rightarrow 0}   \langle \eta_{  ({\bf A, E}) }  \vert U_{\chi} \vert \eta_{  ({\bf A, E}) }  \rangle_\ca = 0,\quad \chi\not= 0
$$
unless $\chi=0$ in which case it equals 1. We note, however, that the expectation value of $U_{\tau_2\chi}$ is non-zero and that the expectation value of  $i \tau_2 E_\chi $ gives
$$
\lim_{\tau_2\rightarrow 0}   \langle \eta_{  ({\bf A, E}) }  \vert i\tau_2 E_\chi \vert \eta_{  ({\bf A, E}) }  \rangle_\ca   =  \int_M \mbox{Tr}_{M_2} {\bf E}(\chi) + \co(\tau_1)
$$
as would be expected. These results combined with the results of the next section show that the question of recovering classical quantities in a classical limit does not appear to be particularly troublesome in this framework -- which is in stark contrast to other approaches to non-perturbative quantum gravity such as loop quantum gravity \cite{Ashtekar:2004eh} (see \cite{Nicolai:2005mc} for an interesting discussion). We expect, however, that the question of a semi-classical limit will, ultimately, involve both parameters $\tau_1$ and $\tau_2$. The latter is a quantization parameter and the former, which appears in the Sobolev norm, is a parameter that can be understood as a dynamical UV regularisation. A semi-classical limit must be local, which implies that $\tau_1\rightarrow0$.

\section{On the dynamics}

Until now we have discussed the representation in $\ch$ of the $\mathbf{QHD}(M)$, which has a natural interpretation in terms of a kinematical sector of a fundamental theory. The task, that remains, is to determine what this theory actually is in terms of physical operators and a dynamical principle. We would like to spend a moment with a general discussion before we go into details.

There are basically two ways to determine a dynamical principle. First, there is the "bottom-up" approach, which is to ask what we would like this theory to deliver in a semi-classical limit and then taylor a dynamical principle accordingly. In the following we shall consider two possible way to do this. The first is to construct operators, which give us the Hamilton of general relativity formulated in terms of Ashtekar variables in a semi-classical limit. Such an operator exist as does an operator, that gives us the Dirac Hamiltonian. Alternative and in a certain sense dual to this is the second possible "bottom-up" approach, which is to try to make contact to the work of Chamseddine and Connes \cite{Connes:1996gi,Chamseddine:2007hz} and in particular to the formulation of the standard model coupled to general relativity in terms of an almost-commutative spectral triple.  The idea here is that since the $\mathbf{HD}(M)$ algebra gives rise to an almost-commutative algebra in a semi-classical limit it is natural to see if structures exist that give rise to a full spectral triple and to see if this might be related to the work of Chamseddine and Connes. As we shall see this too appears to be a viable strategy.

As an alternative to these "bottom-up" approaches it is important to consider also a "top-down" approach, where we seek out structures, which are natural to the kinematical sector generated by the $\mathbf{QHD}(M)$ algebra. Here we see two possibilities. The first is to construct a geometrical structure {\it over} the configuration space $\ca$. The structure, that we have in mind, is a spectral triple type construction that involves the $\mathbf{HD}(M)$ algebra and with a Dirac-type operator that involves the infinitesimal $E_\omega$ operators. This idea was first proposed in \cite{Aastrup:2012vq} and further pursued in   \cite{Aastrup:2014ppa,Aastrup:2015gba,Aastrup:2016ytt,Aastrup:2016caz}. In the following we will analyse this idea in the present framework and we will show that such a structure generates physical operators.

We suspect, however, that a dynamical principle should be obtained by an application of Tomita-Takesaki theory, that says that there exist a unique time flow (up to inner automorphisms) when you have a von Neumann algebra with a cyclic, separating vector. An application of Tomita-Takesaki theory would require us to change our framework to involve Hilbert-Schmidt operators over $M$.

\subsection{A spectral triple on $\ca$}

In the following we construct a spectral triple type construction over the $\mathbf{HD}(M)$ algebra and show that such a metric structure will naturally give rise to key physical operators. \\

We begin by defining an infinite dimensional Clifford algebra $Cl(T^*\ca)$ via the relation
\begin{equation}
\{ \tilde{\xi}_i ,  \tilde{\xi}_j    \} =  -2\a_i \d_{ij},
\label{game}
\end{equation}
where $\{\a_i\}_{i\in\mathbb{N}}$ is a series of real constants, which will be discussed in the following, and where $\{\cdot,\cdot\}$ is the anti-commutator. Note that relation (\ref{game}) implies that we have chosen an orthonormal basis of $\OO^1(M,\mathfrak{su}(2))$, else the right hand side of (\ref{game}) should involve a metric.

The Clifford algebra $Cl(T^*\ca) $ should be properly understood as an inductive limit 
$$
Cl(T^*\ca) :=\lim_{\rightarrow } Cl(T^*\ca_n)
$$
and correspondingly we also construct the modified Hilbert space
$$
L^2(\ca, Cl(T^*\ca)) := \lim_{\rightarrow} L^2(\ca_n, Cl(T^*\ca_n) )
$$
as well as
$$
\ch'= L^2(\ca, Cl(T^*\ca))\otimes L^2(M,S).
$$
Next we construct a Dirac type operator
$$
D_\ca = \sum_i \tilde{\xi}_i \cdot  i \tau_2 E_{\xi_i},
$$
that acts in $L^2(\ca, Cl(T^*\ca)) $. Here the '$\cdot$' should be understood as Clifford multiplication. 
The Dirac type operator $D_\ca$ is defined as a family of operators $D_{\ca_n}$, where each operator $D_{\ca_n}$ is a finite-dimensional Dirac operator acting in $L^2(\ca_n, Cl(T^*\ca_n) )$.\\

In order to get an idea about the correct values of the parameters $\a_i$ let us compute the expectation value of two $E_{\x_i}$ operators on the ground state (\ref{skudiskoven})
\begin{eqnarray}
\left\langle \eta_{  ({\bf A, E}) }  \vert   (i \tau_2)^2 E_{\xi_i}   E_{\xi_j}    \vert \eta_{  ({\bf A, E}) }  \right\rangle_\ca   
&=& - (\tau_2)^2 \frac{d}{ds}\frac{d}{dt}  \left\langle \eta_{  ({\bf A, E}) }  \vert   U_{ t \xi_i + s\xi_j}    \vert \eta_{  ({\bf A, E}) }  \right\rangle_\ca\Big\vert_{s=t=0}
\nn\\
&=&
\frac{\tau_2}{2} \left\langle \xi_i \vert \xi_j \right\rangle_{\mbox{\tiny sob}}  +   \left\langle {\bf e} \vert \xi_i \right\rangle_{\mbox{\tiny sob}} \left\langle {\bf e} \vert \xi_j \right\rangle_{\mbox{\tiny sob}}
\nn\\
&=&
\frac{\tau_2}{2}\d_{ij}  +   b_i b_j
\label{peace}
\end{eqnarray}
where we wrote ${\bf e}=\sum b_i \xi_i$. Now, to have the expectation value of $D_\ca^2$ converge on the ground state we must require
\begin{equation}
\sum_i \a_i < \infty.
\label{louisiana}
\end{equation}

The Dirac type operator $D_\ca$ will not be a Dirac operator in the sense that its resolvent will not be compact, which implies that we do not have a spectral triple in a strict sense. For this reason we refer to $(\mathbf{HD}(M),D_\ca,\ch')$ as a spectral triple type construction; we shall nevertheless use elements of non-commutative in what comes next.\\


Consider first a general spectral triple $(\cb,D,H)$, where $\cb$ is a $C^*$-algebra represented in the Hilbert space $H$ where the self-adjoint Dirac operator $D$ also acts. In the vocabulary of non-commutative geometry a one-form is understood as an object of the form
$$
A= \sum_i a_i [D,b_i],\quad a_i,b_i\in \cb
$$
which gives rise to a so-called {\it fluctuated} Dirac operator of the form 
$$
\tilde{D} = D +A.
$$
In the case of the non-commutative formulation of the standard model such inner fluctuations give rise to the entire bosonic sector of the standard model \cite{Connes:1996gi,Chamseddine:2007hz}.

We are now going to consider inner fluctuations of the Dirac type operator $D_\ca$ 
\begin{equation}
\tilde{D}_\ca = D_\ca + \OO
\label{hjerte}
\end{equation}
where $\OO$ is of the form
$$
\OO = \sum_{i,j} h_i [D_\ca, h_j]
$$
with $h_i\in\mathbf{HD}(M)$. In particular we will consider fluctuations, where the elements in $\mathbf{HD}(M)$ are infinitesimals of the form (\ref{covvar}), i.e. with
$$
\OO=\sum_{i} {\nabla}_{X_i} [D_\ca, {\nabla}_{Y_i}]
$$
where $X_i$ and $Y_i$ are vector fields on $M$. If we let $(x_1,x_2,x_3)$ be a coordinate system on $M$ and set $X_\m= Y_\m=\pa_\m$ for $\m\in \{1,2,3\}$ 
and let\footnote{We use the standard notation with summation over repeated covariant and contra-variant spatial indices. }
$$
\OO= g^{\m\n} {\nabla}_{\pa_\m} [D_\ca, {\nabla}_{\pa_\n}]
$$
then we find 
\begin{equation}
\tilde{D}_\ca = \sum_i  \tilde{\xi}_i \cdot \left( i\tau_2 E_{\xi_i} + i \tau_2  g^{\m\n}   \xi_i ({\pa_\m})  \nabla_{\pa_\n} \right).
\label{fluc}
\end{equation}
Note that this operator has the overall structure
$
\tilde{D}_\ca = D_\ca + D_M
$
with one component acting on $\ca$ and another component acting on $M$. Note also that it depends on the background metric $g$.



\subsection{On the emergence of fermionic QFT}

Consider again the fluctuated Dirac operator (\ref{fluc}) and in particular the one-form over $\ca$
$$
\OO= g^{\m\n}  {\nabla}_{\pa_\m} [D_\ca, {\nabla}_{\pa_\n}]=   \sum_{i}    \tilde{\xi}_i \cdot  i \tau_2 g^{\m\n}   \xi_i ({\pa_\m})  \nabla_{\pa_\n} 
$$
from the previous section. Let us also consider the following operator
\begin{equation}
\Xi =  \OO D_\ca =  g^{\m\n}      {\nabla}_{\pa_\m} [D_\ca, {\nabla}_{\pa_\n}]  D_\ca .
\label{hjerter}
\end{equation}
If we take the trace over the Clifford algebra
$$
\frac{1}{i\tau_2}  \mbox{Tr}_{Cl} \left( \Xi \right) = - 2\sum_{i} i \tau_2    \a_{i} g^{\m\n}    \xi_i ({\pa_\m})  \nabla_{\pa_\n}  E_{\xi_i} 
$$
and compute the expectation value hereof on the ground state (\ref{skudiskoven}), then we get
\begin{eqnarray}
\frac{1}{i\tau_2} \left\langle \eta_{  ({\bf A, E}) }   \vert  \mbox{Tr}_{Cl} \left(  \Xi \right)  \vert\eta_{  ({\bf A, E}) } \right\rangle_{\ca}  
\hspace{-5cm}&&
\nn\\
 &=&
  \lim_{n\rightarrow\infty} \sum_{i} \frac{ 4   \a_{i} }{\left(\sqrt{\p \tau_2}\right)^n}g^{\m\n} \int_{\mathbf{R}^n} (dx)^n  b_i  \xi_i ({\pa_\m}) (\pa_\n  + {\bf A}_\n   )       e^{(-\sum_l\frac{x_l^2}{\tau_2})}
  \nn\\
  &&+
  \lim_{n\rightarrow\infty} \sum_{i} \frac{ 4 i  \a_{i} }{\left(\sqrt{\p \tau_2}\right)^n} g^{\m\n}   \int_{\mathbf{R}^n} (dx)^n   \xi_i ({\pa_\m}) (\sum_{k=1}^n x_k x_i \xi_k(\pa_\n)  \d_{ik} )       e^{(-\sum_l\frac{x_l^2}{\tau_2})}
  \nn
  \end{eqnarray}
 The interpretation of the sum 
 $$
 \sum_{i} g^{\m\n} \a_i b_i \xi_i ({\pa_\m})
 $$
 in the first term depends on how we interpret the parametrisation $\{b_i\}$ as we discussed in the previous section. If we set ${\bf e}=\sum_i b_i\xi_i$ where ${\bf e}=g({\bf E},\cdot)$ is a $\mathfrak{su}(2)$-valued one-form obtained from an inverse triad field ${\bf E}$, then we have
  \begin{equation}
 \sum_{i} g^{\m\n} b_i  \xi_i ({\pa_\m}) = {\bf E}(dx^\n).  
  \label{tower}
  \end{equation}
  We must, however, also include the parameters $\a_i$, and hence we define the modified inverse triad field
  \begin{equation}
   \sum_i g^{\m\n} \a_i b_i \xi_i(\pa_\m) = \tilde{\bf E}(dx^\n).
  \label{iran}
  \end{equation}
  Should we instead have chosen the second interpretation of the parameters $b_i$ according to (\ref{thulesendahl}) then the formulas would look essentially the same except that we would have a correction term at order $\tau_1$. 
  
  With this we continue
   \begin{eqnarray}
   \frac{1}{i\tau_2} \left\langle \eta_{  ({\bf A, E}) }   \vert  \mbox{Tr}_{Cl} \left(  \Xi \right)  \vert\eta_{  ({\bf A, E}) } \right\rangle_{\ca}  
\hspace{-4cm}&&
\nn\\
%
%
%
%
 &=&
  4   \tilde{\bf E}({dx^\m}) (\pa_\m  + {\bf A}_\m   )       
  +
  \sum_{i}  4 i \tau_2  \a_{i}  g^{\m\n}  \xi_i ({\pa_\m})  \xi_i(\pa_\m)      .
\end{eqnarray}
When we write down the corresponding expectation value in $\ch'$ we get
\begin{eqnarray}
\frac{1}{4i\tau_2} \left\langle \eta_{  ({\bf A, E}) } \otimes \psi   \vert   \Xi  \vert\eta_{  ({\bf A, E}) } \otimes \psi \right\rangle_{\ch'}  
\hspace{-5cm}&&\nn\\
&=&
\int_M  \overline{\psi} \sigma^i \tilde{\bf E}^\m_i (\pa_\m  + {\bf A}_\m   )       \psi
  +
 i \tau_2 \sum_{i,\m}   \a_{i}   \int_M  g^{\m\n} \overline{\psi}    \xi_i ({\pa_\m})  \xi_i(\pa_\n)   \psi
 \label{berlingske}
\end{eqnarray}
Here the first term is a spatial Dirac operator evaluated on a spinor on $M$. Note that this term depends on the background metric $g$ via the modification (\ref{iran}). 
The second term in (\ref{berlingske}) is finite due to both (\ref{louisiana}) and (\ref{houston}).
Note that this term too is background dependent. We shall discuss this in section \ref{background}.

We can also include the lapse and shift fields, which in the classical setup encode the foliation of the four-dimensional manifold into a spatial and a temporal part. If ${\bf M}$ is a function that takes values two-by-two matrices
$$
{\bf M}(x) = N(x) \mathrm{1} + N^a(x) \sigma^a
$$
then the operator ${\bf M}\Xi $ will deliver the principal part of the Dirac Hamiltonian in a classical limit where $N(x)$ is then the lapse field and $N^a(x)$ the shift field when contracted with the metric $\bf E$.   \\

Equation (\ref{berlingske}) shows that the spatial Dirac operator turns up in the expectation value of an operator derived from the Dirac type operator $\cd_\ca$. We would like to spend a moment with a possible interpretation of this result.



With a spectral triple type construction $(\mathbf{HD}(M), D_\ca,\ch' )$ we can build an infinite-dimensional exterior algebra generated by $n$-forms over the space $\ca$. The idea, that we propose, is that since the Sobolev norm converges to an $L^2$-norm in the limit $\tau_1\rightarrow 0$ then this exterior algebra and the infinite-dimensional Clifford algebra $Cl(T^*\ca)$ will be linked to the Fock space of a fermionic quantum field theory and that the expectation values of $D_\ca$ on states, which involve $n$-forms, will correspond to $n$-particle states in this Fock space. This idea was first proposed in \cite{Aastrup:2011dt}, where we analysed it in a framework based on a projective system of lattices. The result, that we have obtained above, where the Dirac type operator $D_\ca$ evaluated on a one-form $\OO$ gives us a spatial Dirac operator and, if we include the lapse and shift fields, the Dirac Hamiltonian, suggests that this idea could be realised within quantum holonomy theory.

If this were the case it would imply that the origin of the CAR algebra in this fermionic quantum field theory is the Clifford algebra $Cl(T^*\ca)$, that encodes quantum gravitational data  -- just as the Clifford algebra $Cl(T^*M)$ encodes geometrical data together with a Dirac operator.

\subsection{On the possibility of a gravitational Hamilton operator}

In the classical framework of canonical gravity formulated in terms of Ashtekar variables the Hamilton and diffeomorphism constraints densities have the form
$$
\ch_0 =  \e^{abc}  {\bf E}_a^\m {\bf E}_b^\n  {\bf F}_{\m\n c} ,\quad        \ch_\m =   {\bf E}_a^\n  {\bf F}_{\m\n }^{\;\; a} 
$$
where ${\bf F}_{\m\n}={\bf F}_{\m\n c}\sigma^c$ is the field strength tensor of the Ashtekar connections and where ${\bf E}_a^\m$ is the densitised inverse triad field. It is natural to consider what operators in the present framework might correspond to these classical entities. 
To address this question we first consider how the combination ${\bf E}^\m_a  {\bf E}^\n_b$ could emerge as an expectation value of an operator that involves $U_\omega$. 
Using (\ref{peace}) we write
\begin{equation}
\left\langle \eta_{  ({\bf A, E}) }  \vert   \sum_{ij}   (i \tau_2)^2 ( \xi_i E_{\xi_i} )(\xi_j  E_{\xi_j})      \vert \eta_{  ({\bf A, E}) }  \right\rangle_\ca   = {\bf e} {\bf e} +i \tau_2 \sum_i   \xi_i \xi_i
\label{olav}
\end{equation}
where ${\bf e}=\sum b_i \xi_i$.
Now, equation (\ref{olav}) is a two-form, which implies that we can let it act on two vectors. We therefore consider the operator\footnote{We here ignore the question of operator ordering.}
$$
 \sum_{ij,\m\n}   (i \tau_2)^2  \xi_i(\pa_\a)  \xi_j(\pa_\b)    \nabla_{\pa_\m} \nabla_{\pa_\n} E_{\xi_i}  E_{\xi_j} 
$$
as well as the expectation value
\begin{eqnarray}
\left\langle \eta_{  ({\bf A, E}) }  \vert   (i \tau_2)^2   \nabla_{\pa_\m} \nabla_{\pa_\n} E_{\xi_i}  E_{\xi_j}       \vert \eta_{  ({\bf A, E}) }  \right\rangle_\ca   
\hspace{-6cm}&&
\nn\\&=&
 \lim_{n\rightarrow\infty} \frac{  (i \tau_2)^2  }{\left(\sqrt{\p \tau_2}\right)^n} \int_{\mathbf{R}^n} (dx)^n  
  \left(\pa_\m  +{\bf A}_\m  +\sum_{k_1=1}^n x_{k_1}  \xi_{k_1} (\pa_\m)  \right) 
   \nn\\&&
\times   \left(\pa_\n  +{\bf A}_\n +\sum_{k_2=1}^n x_{k_2} \xi_{k_2} (\pa_\n)  \right) \left(  \left(\frac{ x_i  - i b_i }{\tau_2}\right)  \left(\frac{ x_j  - i b_j }{\tau_2}\right)  + \frac{\d_{ij}}{\tau_2}   \right)   e^{(-\sum_l\frac{x_l^2}{\tau_2})}
  \nn\\
  &=& { \nabla}^{\mbox{\tiny cl}}_\m { \nabla}^{\mbox{\tiny cl}}_\n  \left( b_i b_j - \frac{3\tau_2}{4} \d_{ij}   \right) 
  + \frac{i \tau_2}{2} { \nabla}^{\mbox{\tiny cl}}_\m  \left(  \xi_{i} (\pa_\n)  b_j  +  \xi_{j} (\pa_\n)  b_i \right)
  \nn\\&&
  +  \frac{i\tau_2}{2}   \left(  \xi_{i} (\pa_\m)  b_j  +  \xi_{j} (\pa_\m)  b_i \right){ \nabla}^{\mbox{\tiny cl}}_\n
    \nn\\&&
  +  \sum_k \xi_{k} (\pa_\m)    \xi_k (\pa_\n)   \left(  \frac{\tau_2}{2} b_i b_j  - \frac{3 \tau_2^2}{4} \d_{ij}  \right)
    - \frac{ \tau_2^2}{2} \xi_{i} (\pa_\m)    \xi_j (\pa_\n)     \d_{ij}  
    \label{skolen}
\end{eqnarray}
where ${ \nabla}^{\mbox{\tiny cl}}_\m = \pa_\m + {\bf A}_\m$. This result show that the operator\footnote{here we ignore questions regarding operator ordering.}
\begin{equation}
{\mathbbmss{H}}_0 :=      (i \tau_2)^2 \sum_{ij,\m\n} g^{\a\m} g^{\b\n} \xi_i(\pa_\a)\xi_j(\pa_\b) [ \nabla_{\pa_\m}, \nabla_{\pa_\n}] E_{\xi_i}  E_{\xi_j}    
\label{sommervind}
\end{equation}
gives the integral over the classical Hamilton density in a classical limit
\begin{equation}
\left\langle \eta_{  ({\bf A, E}) } \otimes \mathrm{1} \vert  {\mathbbmss{H}}_0     \vert \eta_{  ({\bf A, E}) }  \otimes \mathrm{1} \right\rangle_{\ch} = \int_M \ch_0 + \co(\tau_2)
\label{lostcityofZ}
\end{equation}
when evaluated on the state $\eta_{  ({\bf A, E}) } \otimes \mathrm{1}$ in $\ch$.

Note again that the expectation value of $ {\mathbbmss{H}}_0$ will depend on the background metric $g$, just as we saw it in the previous section with the spatial Dirac operator and Dirac Hamiltonian. We shall discuss this background dependency in section \ref{background}.

Finally, as we discussed above for the Dirac Hamiltonian we can also include the lapse and shift fields by introducing a function ${\bf M}$ that takes values two-by-two matrices. Then the operator ${\bf M} {\mathbbmss{H}}_0$ will deliver both the Hamilton and Diffeomorphism constraints in a classical limit.   \\



\subsection{Density weight of the triad field}
\label{density}

Before we move on to discuss background dependency we would like to pause for a moment to discuss the classical point ${\bf E}$ introduced in (\ref{skudiskoven}).
It is an interesting question whether or not ${\bf E}$ should be understood as a {\it densitised} inverse triad field, i.e. whether it includes the square root of the determinant of the metric, denoted by $e$, or not. If ${\bf E}$ is not densitised then the metric density must come from the inner product in the Hilbert space $\ch$, a possibility that we discussed in \cite{Aastrup:2016ytt}. This would mean that the density $e$ is not a part of the quantised data but comes as a classical 'ad-on', a significant departure from the framework of canonical quantum gravity. We can of course interpret ${\bf E}$ to be a densitised field -- and we believe that this is the natural course of action -- but  this choice makes the formulation of the Hamilton constraint operator ${\mathbbmss{H}}_0  $ in (\ref{lostcityofZ}) more difficult.

The point is that if we choose ${\bf E}$ to be a field of density weight one, that is, the Ashtekar triad field, then we need to divide the operator $ {\mathbbmss{H}}_0$ with $e$ in order to obtain the right density weight. 
What we encounter here is a problem, that has been discussed extensively within the framework of loop quantum gravity (see for example \cite{Ashtekar:2004eh}). We do, however, not believe that the solutions discussed there are applicable nor natural in the present framework. Rather, we interpret the emergence of this issue as a hint that the introduction of an operator like (\ref{sommervind}) is not the right approach but that we should instead focus on a link to the framework of non-commutative geometry. We shall discuss this further in section \ref{NCG}.

\section{On background dependency}
\label{background}

Let us now discuss the question of background dependency. 
The metric dependency shows up in essentially four different ways:
\begin{enumerate}
\item
First of all, the Hodge-Laplace operator in the Sobolev norm (\ref{sob}) depends on the background metric $g$.
\item
Second, the norm (\ref{sob}), even if we turn off the term with the Hodge-Laplace operator, depends on the background metric in its $L^2$-sector, since this requires an inner product on $\OO^1(M,\mathfrak{su}(2))$.
\item
Third, the inner product on $L^2(M,S)$ requires a metric.
\item
Fourth, a 3-metric appears in the phase (\ref{skudiskoven}) that gives rise to an inverse triad field. 
\end{enumerate}
Of these four metric dependencies the first three would normally be thought of as an actual background dependency (the fourth being related to a semi-classical limit). Note, of course, that the Sobolev norm (\ref{sob}) is absolutely critical for our construction, without it the representation of the $\mathbf{QHD}(M)$ algebra does not exist. Note also that the metric referred to in point 4. need not coincide with the metric referred to in point 1-3. Finally, the metric density required for the inner product on $L^2(M,S)$ could come from the inverse triad field in (\ref{skudiskoven}) if this has density weight one as we discussed in section \ref{density}.

One effect of this metric dependency, that we have already seen, is that the expectation values of physical operators related to the Dirac and gravitational Hamiltonians in (\ref{berlingske}) and (\ref{lostcityofZ}) depend both on the background metric from the Hodge-Laplace operator and from the metric originating from the phase in (\ref{skudiskoven}).\\


The key to the discussion of background metric dependency is the map $\rho_g: \ca \rightarrow \mathds{R}^\infty$ 
$$
 \rho_g(\omega) = (x_1,x_2,\ldots ) 
$$
for $ \omega= \sum_i x_i \xi_i \in \ca$, that embeds the configuration space $\ca$ into the projective limit $\mathds{R}^\infty$ and thereby gives us a representation. These representations are labelled by the metric $g$ as the Sobolev eigenvectors $\{\xi_i\}$ are metric dependent. Now, the question is whether some of these representations are unitary equivalent. To discuss this consider first the state $\eta_{\bf A}$ in (\ref{groundstate}) as well as the state $\eta'_{\bf A}$, which we obtain from $\eta_{\bf A}$ by modifying each Gaussian factor by
$$
e^{-\frac{(x_{i}-a_i)^2}{2\tau_2}} \longrightarrow e^{-\frac{(x_{i}-a_i)^2}{2 \k \tau_2}}\quad \forall  i\in\mathbb{N}
$$
where $\k$ is a real number. We then have
\begin{equation}
\langle  \eta_{\bf A}\vert \eta_{\bf A}' \rangle_\ca =
\left\{
\begin{array}{ll}
0& \mbox{for}\quad k\not= 0\\
1& \mbox{for}\quad k= 0
\end{array}
\right. .
\label{vangelis}
\end{equation}
This family of ground states $\{\eta_{\bf A}(\k)\}_{\k\in \mathbb{R}_+}$ corresponds to different metrics $g$ and $g'$ and equation (\ref{vangelis}) then tells us that they are all orthogonal. In fact, it can be shown that the entire representations given by $g$ and $g'$ will be orthogonal. We believe that this implies that these representations cannot be unitarily equivalent.

The question remains, however, if {\it any} representations given by different metrics are unitarily equivalent. We hope to be able to answer this question in the future.\\

Nevertheless, the picture, that is emerging, is that whereas the $\mathbf{QHD}(M)$ algebra is metric independent its representations are metric dependent. That this dependency is physical seems clear when we compute for instance the expectation value in (\ref{berlingske}). We believe that this situation invites two possible interpretations. 
\begin{enumerate}
\item[1)]
The first interpretation is that each representation of the $\mathbf{QHD}(M)$ algebra should be understood in terms of a semi-classical phase. Note that there exist also a highly symmetric state
$$
\rho_{\mbox{\tiny sym}} (f e^X) = 0,\quad \rho_{\mbox{\tiny sym}}(U_\omega) =1,
$$
that gives rise to a very simple representation.
One might therefore speculate that we have a scenario akin to the Ising model where there is both a highly symmetric phase as well as a phase, that corresponds to a broken symmetry, which in our case would correspond to a choice of a background metric.
\item[2)]
The second interpretation is that this framework should be interpreted in terms of a quantum Yang-Mills theory. The kinematical sector of a $SU(2)$ Yang-Mills theory is identical to that of gravity formulated in terms of Ashtekar variables and the framework that we have presented so far can be straight forwardly generalised to arbitrary compact Lie groups and to compact manifolds of any dimension.

Moreover, the framework can be generalised to other field theories, such as scalar theories for instance. 
We shall discuss this second interpretation in more detail in the next section.

\end{enumerate}

\section{Non-perturbative quantum field theory}

In this section we will discuss the application of the framework, that we have presented so far, to general field theories such as Yang-Mills and scalar theories.

\subsection{Yang-Mills theory }

In the Hamiltonian formulation of Yang-Mills theory in 3+1 dimensions the Hamilton density has the form \cite{Jackiw:1979ur} 
\begin{equation}
\ch_{\mbox{\tiny YM}} = E^2 + B^2
\label{borebore}
\end{equation}
where $E^\m_a=F^{\m0}_a$ and where $B_a^\m=-\frac{1}{2}\e^{\m\a\b}F_{a\a\b}$ with $F$ being the field strength tensor and the index $a$ takes values in the Lie algebra of a compact gauge group $G$ (the indices $\m,\n,...$ are spatial indices). The non-vanishing canonical commutation relations are identical to (\ref{fix}) when these are written for a general gauge group.

As we have already pointed out this means that the kinematical sector given by the representation of the $\mathbf{QHD}(M)$ algebra in $\ch$, which is straight forwardly generalised to an arbitrary compact group $G$, is identical to that of a Yang-Mills theory. 
The Yang-Mills Hamilton operator then has the form 
$$
{\mathbbmss{H}}_{\mbox{\tiny YM}}  = {\mathbbmss{H}}_{1}  + {\mathbbmss{H}}_{2}  
$$
with 
\begin{eqnarray}
{\mathbbmss{H}}_{1}  &=& (i \tau_2)^2 \sum_{ij} g^{\m\n}  \xi_i(\pa_\m)   \xi_j(\pa_\n)  E_{\xi_i} E_{\xi_j} 
\nn\\
{\mathbbmss{H}}_{2}  &=& \frac{1}{2}  g^{\a\m} g^{\b\n}  [\nabla_{\pa_\m} , \nabla_{\pa_\n} ]  [\nabla_{\pa_\a} , \nabla_{\pa_\b} ] 
\end{eqnarray}

We are now going to consider the computation of the expectation value of ${\mathbbmss{H}}_{\mbox{\tiny YM}} $ on the ground state $\eta_{({\bf A,E})}$. Let us first take ${\mathbbmss{H}}_{1} $, where we can use the result in (\ref{peace}) to write
\begin{eqnarray}
\big\langle \eta_{  ({\bf A, E}) }  \big\vert  {\mathbbmss{H}}_{1}    \big\vert \eta_{  ({\bf A, E}) }  \big\rangle_\ca   
&=&
 \sum_{ij}  g^{\m\n}  \xi_i(\pa_\m)   \xi_j(\pa_\n) \left(  \frac{\tau_2}{2}\d_{ij}  +   b_i b_j  \right)
 \nn\\
 &=& g_{\m\n} {\bf E}(dx^\m) {\bf E}(dx^\n) 
 +  \frac{\tau_2}{2} \sum_i  g^{\m\n}  \xi_i(\pa_\m)   \xi_i(\pa_\n).
\label{Hannoverbynight}
\end{eqnarray}
Here ${\bf E}$ is a vector field according to (\ref{tower}) except that it now takes values in the Lie algebra of $G$ instead of $\mathfrak{su}(2)$.

The first term in (\ref{Hannoverbynight}) will give us the first term $E^2$ of the classical Hamiltonian (\ref{borebore}). 
The second term in  (\ref{Hannoverbynight}) is pure 'quantum'. The condition (\ref{houston}) ensures that it is finite but it diverges in the limit $\tau_1\rightarrow 0$.

Next, to see what the expectation value of ${\mathbbmss{H}}_{2} $ looks like we first compute
\begin{eqnarray}
\big\langle \eta_{  ({\bf A, E}) }  \big\vert   \nabla_{\pa_{\m_1}} \nabla_{\pa_{\m_2}}  \nabla_{\pa_{\m_3}}  \nabla_{\pa_{\m_4}}    \big\vert \eta_{  ({\bf A, E}) }  \big\rangle_\ca   
\hspace{-6cm} &&\nn\\
&=&
{ \nabla}^{\mbox{\tiny cl}}_{\m_1}{ \nabla}^{\mbox{\tiny cl}}_{\m_2}{ \nabla}^{\mbox{\tiny cl}}_{\m_3}{ \nabla}^{\mbox{\tiny cl}}_{\m_4} 
\nn\\&&
+   \frac{\tau_2}{2}   { \nabla}^{\mbox{\tiny cl}}_{\m_1}{ \nabla}^{\mbox{\tiny cl}}_{\m_2} \sum_k  \xi_{k} (\pa_{\m_3}) \xi_{k} (\pa_{\m_4}) + \mbox{5 perm.} (\nabla_\pa,\nabla_\pa,  \xi_{k} (\pa), \xi_{k} (\pa))
\nn\\
&&+  \frac{\tau_2^2}{4} \sum_{k_1 k_2}  \Big(  \xi_{k_1} (\pa_{\m_1}) \xi_{k_1} (\pa_{\m_2}) \xi_{k_2} (\pa_{\m_3}) \xi_{k_2} (\pa_{\m_4})+
\nn\\&&
       \xi_{k_1} (\pa_{\m_1}) \xi_{k_2} (\pa_{\m_2}) \xi_{k_1} (\pa_{\m_3}) \xi_{k_2} (\pa_{\m_4})   +   \xi_{k_1} (\pa_{\m_1}) \xi_{k_2} (\pa_{\m_2}) \xi_{k_2} (\pa_{\m_3}) \xi_{k_1} (\pa_{\m_4})   \Big)
       \nn\\&&
       + \frac{\tau_2^2}{2} \sum_k  \xi_{k} (\pa_{\m_1}) \xi_{k} (\pa_{\m_2}) \xi_{k} (\pa_{\m_3}) \xi_{k} (\pa_{\m_4})  
       \label{rain}
\end{eqnarray}
where we again write ${ \nabla}^{\mbox{\tiny cl}}_\m = \pa_\m + {\bf A}_\m$. Again we can use condition (\ref{houston}) to ensure that the various sums over contracted Sobolev eigenvectors $\xi_i$ in (\ref{rain}) are finite. Note again that they diverge in the limit $\tau_1\rightarrow 0$.

The computation (\ref{rain}) shows that the expectation value of ${\mathbbmss{H}}_{2} $ will be of the form
\begin{eqnarray}
\big\langle \eta_{  ({\bf A, E}) }  \big\vert  {\mathbbmss{H}}_{2}   \big\vert \eta_{  ({\bf A, E}) }  \big\rangle_\ca   
=
 \frac{1}{2}  g^{\a\m} g^{\b\n}  [\nabla^{\mbox{\tiny cl}}_{\m} , \nabla^{\mbox{\tiny cl}}_{\n} ]  [\nabla^{\mbox{\tiny cl}}_{\a} , \nabla^{\mbox{\tiny cl}}_{\b} ]  +  \tau_2  {\mathbbmss{H}}_{2}^q
\end{eqnarray}
where we note that the quantum correction $ {\mathbbmss{H}}_{2}^q$ a priori does not vanishes if $G$ is Abelian, as we might have expected. The second term in (\ref{Hannoverbynight}) also does not vanish in the Abelian case. If we write $\Delta =\int_M \frac{\tau_2}{4} \sum_i  g^{\m\n}  \xi_i(\pa_\m)   \xi_i(\pa_\n)$ then we have
\begin{equation}
0< \Delta  <\infty.
\label{berlin}
\end{equation}
There will also be a contribution from $ {\mathbbmss{H}}_{2} $, which we have not written down.  Note that $\Delta$ diverges in the limit $\tau_1\rightarrow 0$. 
Note also that $\Delta$ is non-zero also in the Abelian case.
One might have expected that a quantum $U(1)$ gauge theory would have no quantum effects, but that does not appear to be the case.\\

What we see here is that the representation of the $\mathbf{QHD}(M)$ algebra provides a framework for a non-perturbative quantum Yang-Mills theory. It is remarkable that this quantum theory is completely well defined. 
It is an interesting question if the framework presented here will coincide with results known from perturbative quantum field theory in an expansion around the double limit $\tau_1\rightarrow 0$ and $\tau_2\rightarrow 0$.

\subsection{Scalar and other field theories}

The framework, that we have presented in this paper can also be generalised to field theories that are not gauge theories. In the following we will first consider scalar fields.

Let now $\{\b_i\}_{i\in\mathbb{N}}$ be an orthonormal basis of $C^\infty(M)$ with respect to a Sobolev norm $\Vert\cdot\Vert_{\mbox{\tiny sob}}$. With this we can again construct a space $L^2(\Phi)$ as a projective limit over intermediate spaces $L^2(\Phi_n)$ with an inner product given by
$$
\langle \eta \vert \zeta \rangle_{\Phi_n} = \int_{\mathbb{R}^n} \overline{\eta(x_1\b_1 + \ldots + x_n \b_n)} \zeta (x_1\b_1 + \ldots + x_n \b_n) dx_1\ldots dx_n,
$$
where $\eta$ and $\zeta$ are elements in $L^2(\Phi)$. The translation operator $U_f$, $f$ being a function on $M$, then act by
$$
U_f \eta(\phi) = \eta(\phi + f)
$$
With scalar fields we do not have an algebra like the $\mathbf{QHD}(M)$ algebra but we can instead take the algebra $C^\infty(M)$. This means that a field acts on $L^2(M)$ by
$$
\phi(f) (m) = \phi(m) f(m) ,\quad f\in L^2(M)
$$
Note that the algebra generated by the $U_f$ operators and $C^\infty(M)$ encodes a smeared version of the canonical commutation relations for scalar fields.

Finally, we build the full Hilbert space
$$
\ch_{\mbox{\tiny scalar}} = L^2(\Phi)\otimes L^2(M)
$$
where we have a representation of an algebra involving both translation operators and field operators.

In this way we can construct various scalar field theories such as $\phi^4$-theory etc.
We will not work out any details but simply note that such a construction with scalar fields based on an expansion over Sobolev eigenvectors is possible and will lead to a non-perturbative quantum field theory. Likewise, we note that similar constructions for any field theory appear to be possible. At the present level of analysis we do not see any obstacles for a general framework for non-perturbative quantum field theory to be constructed in this way.


\section{Big bang and black hole singularities 
}
\label{blackholes}

A key feature of the Hilbert space representation of the $\mathbf{QHD}(M)$ algebra is that it is non-local. The bulk of the measure in $L^2(\ca)$ consist of connections, which are differentiable\footnote{This statement depends on the choice of $\sigma$ in (\ref{sob}), see \cite{AAA1}.} \cite{AAA1}, and in general field configurations are weighted according to their Sobolev norm. 
Since the Sobolev norm dominates the supremum norm \cite{Nirenberg} this implies that field configurations {cannot} have singular points and thus we conclude that states in $L^2(\ca)$ and $\ch$ {cannot} include geometries, which are singular. Also, if we consider expectation values of the infinitesimal operators $E_\omega$, which corresponds classically to the triad field, then we find that transitions between field configurations, that are respectively very smooth and very far from smooth are suppressed. In particular are transition from regular to singular field configurations assigned zero probability. This suggest, once more, that singular field configurations cannot occur in the present framework.

These argumentations appear to imply that singularities such as the initial big bang singularity and the singularities purported to reside at the centre of black holes cannot exist within the framework of quantum holonomy theory.


A related issue is that of the large scale structure of the observable Universe. At present the preferred explanation for the near-flatness of the Universe involves inflation theory but one might speculate whether a framework that involves quantum holonomy theory could provide an alternative explanation. Indeed, since transitions between different field configurations are biased towards smooth geometries it seems possible that this could provide a mechanism, where an early Universe, that is dominated by quantum effects of the gravitational field, is more likely to evolve into a near-flat Universe than otherwise. 


\section{An emergent almost-commutative spectral triple}
\label{NCG}


As we have pointed out in \cite{Aastrup:2012vq} a sub-algebra of the $\mathbf{HD}(M)$ algebra gives rise to an almost-commutative algebra in a classical limit. This opens the door to a possible connection to the standard model of particle physics via Chamseddine and Connes work \cite{Connes:1996gi,Chamseddine:2007hz}.
To see this let us start by noting that the algebra $\mathbf{HD}(M)$ can be formulated as the closure of the semi-direct product
\begin{equation}
\mathbf{HD}(M)  = C_c^\infty(M)  \rtimes  \cF  / I. 
\label{alg1}
\end{equation}
in the norms described in \cite{AGnew}, where $\cf$ is the group generated by operators $e^X$, where $I$ is an ideal given by certain reparametrizations of flows (see \cite{AGnew} for details) and where $C^\infty_c (M) $ is the algebra of smooth functions with compact support. The semi-direct product comes with the multiplication relation
$$f_1F_1 f_2 F_2=f_1 F_1 (f_2) F_1 F_2 \;,  $$
where $F_1,F_2\in\cf$.
In \cite{Aastrup:2012vq} we observed that $\mathbf{HD}(M)$ reduces in a classical limit to the algebra
\begin{equation}
 \left(C_c^\infty(M)\otimes M_2(\mathbb{C})\right) 
\label{almost}
\end{equation}
if we restrict it to closed flows.
This is so because the holonomies on a fixed classical geometry generate a two-by-two matrix algebra\footnote{We assume we are considering a semi-classical analysis around a irreducible connection.}.
Thus we find an almost-commutative algebra in a semi-classical limit when we consider closed flows.

Now, a Dirac operator, that emerges from quantum holonomy theory will necessarily interact {\it both} with the algebra $C_c^\infty(M)$ of functions on $M$ and with the finite dimensional matrix algebra $M_2(\mathbb{C}) $, that emerges from the holonomies -- which is precisely what one would expect in order to se an almost-commutative {\it spectral triple} emerge.  
Consider first the fluctuated Dirac operator in (\ref{fluc}). Here it is clear that this operator has both a spatial and what will be a finite part in a semi-classical limit, given by the fluctuation involving $g^{\m\n}\xi_i(\pa_\n)\nabla_\m$ and by the vector field $E_{\xi_i}$ respectively. This operator involves, however, the Clifford algebra over $\ca$ and one would need more analysis to see how this algebra might be interpreted. 
Alternatively,  consider the operator $\Xi$ in (\ref{hjerter}), which on the one hand gives rise to the spatial Dirac operator in a classical limit and on the other hand has again a non-trivial interaction with the holonomies via the operators $E_{\xi_i}$. 

Regardless of which operator we consider it is clear that the general structure, that emerges, is that of an almost-commutative spectral triple -- namely a Dirac operator, that interacts non-trivially with both parts of $C_c^\infty(M)\otimes M_2(\mathbb{C}) $. It is an interesting question whether this could be related to the work of Chamseddine and Connes on the standard model \cite{Connes:1996gi,Chamseddine:2007hz}.

\section{Discussion and outlook}
\label{discussion}

In this paper and its companion \cite{AAA1} we have established the mathematical existence of a non-perturbative theory of quantum gravity known as quantum holonomy theory by proving that a Hilbert space representation of the $\mathbf{QHD}(M)$ algebra exist. We have shown that operators exist, which gives the Hamiltonians of general relativity and matter couplings when evaluated on certain states and in a classical limit -- and in addition to this we have argued that the structure of an almost-commutative spectral triple emerges in the same limit. These results opens a door to a new theoretical framework of quantum gravity with a potential connection to the standard model of particle physics. \\

Probably the most pressing open problem now is to determine which representations of the $\mathbf{QHD}(M)$ algebra are unitary equivalent. Since each representation is labelled by a 3-metric this issue is intimately related to the question of background dependency. It seems clear that most geometries will give rise to inequivalent representations because they depend on a scale. What is less clear is whether any representations are unitarily equivalent. This issue is also related to the choice of exponent of the Hodge-Laplace operator  in the Sobolev norm -- this too labels different representations -- and the support of the Hilbert space measure, where higher exponents shifts the support towards smoother one-forms.

Note also that two metrics, that differ only at scales far below the Planck scale, may still be unitarily in-equivalent. This raises the question what physical impact the infinite limit, that corresponds to extremely small scales, will have.

What remains, however, is that the theory we have presented is background dependent. One possible interpretation of this dependency is that the representations, which we have found, corresponds to a broken phase, i.e. a choice of a metric, in a theory akin to the Ising model, where the quantum theory then gives us fluctuations around this metric. \\

Our aim with the construction, that we have presented, is to find a theory of quantum gravity and in particular to explain the origin of the almost-commutative spectral triple, that Chamseddine and Connes have shown to be related to the standard model of particle physics. Although we find strong evidence that the structure of an almost-commutative spectral triple will emerge in a local and semi-classical limit it still remains to be proven. What is missing is a clear understanding of the dynamics of the theory. We have proposed to construct a Dirac type operator and a spectral triple -- that is, a metric structure {\it over} the configuration space of connections -- but more work is required to fully understand how such an operator might generate a time-flow and physical quantities. A more substantial comparison with the work of Chamseddine and Connes will not be possible until these issues are settled.

The idea, that we are pursuing, is that a quantum theory of gravity will produce an almost-commutative spectral triple in a local and semi-classical limit and that perturbations around this limit will turn out to coincide with a quantum field theory of the gauge and Higgs degrees of freedom generated by inner automorphisms in this triple -- the bosonic sector of the standard model -- as well as the fermionic degrees of freedom. For the latter we suspect that the CAR algebra will come from the infinite-dimensional Clifford algebra used to define the Dirac type operator over the configuration space of connections. Whether this idea can be realised remains, of course, open, but we believe that the results presented in this paper represents a substantial step in the right direction.\\

The question of formulating a rigorous framework of non-perturbative quantum field theory is one of the most critical unsolved problems in modern theoretical physics. Since the framework, that we have presented, appears to offer a solution to this problem it is interesting to study also other field theories than gravity within this framework and to determine whether such quantum theories coincides with perturbative quantum field theory in a local and semi-classical limit. Here it will be interesting to see what distinguishes renormalizable and non-renormalizable theories and how renormalisation theory might emerge. It is clear that quantum corrections will diverge in the local limit $\tau_1\rightarrow 0$ and it seems reasonable to expect that renormalisation theory emerges as a way to expand around the local and semi-classical limit. 
Another interesting question is whether this framework could be used to study non-perturbative effects in quantum field theory. Here we have already computed what appears to be a mass gap in Yang-Mills theory, a result that requires further analysis and scrutiny.

The reason why our framework succeeds where ordinary quantum field theory does not, is that we do not have local field operators. The non-locality introduced by the Sobolev norm is the crucial ingredient that introduces a UV dampening effect that permits a representation of an algebra that encodes the canonical commutation relations. Note that the trick with the Sobolev norm is not exclusive for the $\mathbf{QHD}(M)$ algebra, it seems straight forwardly applicable to other field theories too.

One intriguing consequence of this non-locality is that quantum transitions between regular and singular field configurations have zero probability in the Hilbert space. With the proviso that we do not yet have a clear understanding of the dynamics, this seems to suggest that the formation of for instance black hole singularities and, possibly, the big bang singularity is impossible within our framework.\\

It is interesting that the gauge symmetry does not play a decisive role in our construction. Since different elements in a gauge orbit will have different Sobolev norm -- the range goes from zero to infinite -- we can immediately conclude that the Hilbert space measure is not invariant under gauge transformations. It is an interesting question what role the gauge symmetry plays -- certainly it makes sense in a quantum theory of gravity that gauge transformations, that vary at a scale far below the Planck scale, are treated differently from transformations, that vary only a large scales. It is possible that the gauge symmetry will only play an important role in a semi-classical and local limit.

Note also that the original Ashtekar connection takes values in the (anti) self-dual sector of $SL(2,\mathbb{C})$. This means that the original Ashtekar connection is complex and that one must introduce a reality condition to identify a physical sector. The connection, that we are using, is, however, a real $SU(2)$ connection, which corresponds to a Euclidean signature. 
It is an interesting question how complex variables and a reality condition might fit into our construction. Here an important clue might be that the basis of Sobolev eigenvectors, that we use to build the Hilbert space representation, must be chosen to be real for our proof to work (whether an alternative proof for a complex basis exist we do not know. We suspect that by choosing the exponent of the Hodge-Laplace operator large enough a proof might exist). This choice of a real basis could be interpreted in terms of a reality condition and thus hint at a framework, that involves complexified connections.\\

As a final remark note that it is a widespread assumption about a theory of quantum gravity that it will involve topology as a variable and that quantum effects of the gravitational field will include changes in topology. The construction, that we have presented, offers a completely different picture, where the topology of the manifold is not a physical variable and where the Hilbert space measure weighs quantum transitions according to their variation -- and where the singular ones are forbidden. We find the possibility of a 'soft' theory of quantum gravity intriguing.

\vspace{1cm}
\noindent{\bf\large Acknowledgements}\\

We would like to thank Professor Yum-Tong Siu for providing a reference on the Sobolev norm. We would also like to thank Professor Daniel Grieser for help on the Hodge-Laplace operator.
JMG would like to express his gratitude towards Ilyas Khan, United Kingdom, for his generous financial support. JMG would also like to express his gratitude towards the following list of sponsors:  Ria Blanken, Niels Peter Dahl, Simon Kitson, Rita and Hans-J\o rgen Mogensen, Tero Pulkkinen and Christopher Skak for their financial support, as well as all the backers of the 2016 Indiegogo crowdfunding campaign, that has enabled this work.

\end{document}